\documentclass[sigconf]{acmart}

\AtBeginDocument{%
  }
	
\usepackage{graphicx}
\usepackage{amsmath,amsfonts}
\usepackage{booktabs} 
\usepackage[10pt]{moresize}
\usepackage{graphicx}
\usepackage[english]{babel}
\usepackage{graphicx}
\usepackage{tcolorbox}
\usepackage{listings}
\usepackage{float}
\usepackage{multirow}
\usepackage[scaled]{helvet}
\usepackage{mathrsfs}
\usepackage[linesnumbered,ruled,lined,noend]{algorithm2e}
\usepackage{mathpartir}
\usepackage{mathtools}
\usepackage{dsfont} 
\usepackage{stmaryrd}
\usepackage{url}
\usepackage{textcomp} 
\usepackage{bbm}
\usepackage{verbdef}
\usepackage{xspace}
\usepackage{verbatim}
\usepackage{lipsum}
\usepackage{wrapfig}

\usepackage{enumitem}
\usepackage{pifont}
\usepackage{varwidth}
\usepackage{xpatch}
\usepackage{multirow}
\usepackage{xcolor}
\usepackage[export]{adjustbox}
\usepackage{caption}
\usepackage{subcaption}
\usepackage[normalem]{ulem}
\usepackage[toc,page]{appendix}

\usepackage[varqu]{zi4}

\usepackage{flushend}

\def\toolname{\textsc{WVProfiler }}

\usepackage{hyperref}
\usepackage{todonotes}
\usepackage{cleveref}
\usepackage{fancyvrb}
\usepackage{listings}

\definecolor{clr-background}{RGB}{255,255,255}
\definecolor{clr-text}{RGB}{0,0,0}
\definecolor{clr-string}{RGB}{163,21,21}
\definecolor{clr-namespace}{RGB}{0,0,0}
\definecolor{clr-preprocessor}{RGB}{128,128,128}
\definecolor{clr-keyword}{RGB}{0,0,255}
\definecolor{clr-type}{RGB}{43,145,175}
\definecolor{clr-variable}{RGB}{0,0,0}
\definecolor{clr-constant}{RGB}{111,0,138} 
\definecolor{clr-comment}{RGB}{0,128,0}

\lstdefinestyle{fingerprint}{
    basicstyle=\scriptsize\ttfamily
}

\lstdefinestyle{VS2017}{
	backgroundcolor=\color{clr-background},
	basicstyle=\color{clr-text}, 
	stringstyle=\color{clr-string},
	identifierstyle=\color{clr-variable}, 
	commentstyle=\color{clr-comment},
	directivestyle=\color{clr-preprocessor}, 
	keywordstyle=\color{clr-type},
	keywordstyle={[2]\color{clr-constant}}, 
	tabsize=2,
}
\lstdefinelanguage{JavaScript}{
	keywords={typeof, new, true, false, catch, function, return, null, catch, switch, var, if, in, while, do, else, case, break},
	keywordstyle=\color{blue}\bfseries,
	ndkeywords={class, export, boolean, throw, implements, import, this},
	ndkeywordstyle=\color{darkgray}\bfseries,
	identifierstyle=\color{black},
	sensitive=false,
	comment=[l]{//},
	morecomment=[s]{/*}{*/},
	commentstyle=\color{purple}\ttfamily,
	stringstyle=\color{red}\ttfamily,
	morestring=[b]',
	morestring=[b]"
}

\let\origthelstnumber\thelstnumber
\makeatletter
\newcommand*\Suppressnumber{%
	\lst@AddToHook{OnNewLine}{%
		\let\thelstnumber\relax%
		\advance\c@lstnumber-\@ne\relax%
	}%
}

\newcommand*\Reactivatenumber{%
	\lst@AddToHook{OnNewLine}{%
		\let\thelstnumber\origthelstnumber%
		\advance\c@lstnumber\@ne\relax}%
}
\theoremstyle{definition}

\begin{document}
	\sloppy
\title{Our fingerprints don't fade from the \textit{Apps} we touch: Fingerprinting the Android WebView}
%
%
\author{Abhishek Tiwari}
\email{abhishek.tiwari@uni-passau.de}
\affiliation{Faculty of Computer Science and Mathematics,\\University of Passau
	\country{Germany}
}
\authornotemark[1]
\author{Jyoti Prakash}
\email{jyoti.prakash@uni-passau.de}
\affiliation{Faculty of Computer Science and Mathematics,\\University of Passau
\country{Germany}
}\authornote{Joint First Author}

\author{Alimerdan Rahimov}
\email{rahimo01@ads.uni-passau.de}
\affiliation{Faculty of Computer Science and Mathematics,\\University of Passau
	\country{Germany}
}
\author{Christian Hammer}
\email{christian.hammer@uni-passau.de}
\affiliation{Faculty of Computer Science and Mathematics,\\University of Passau
\country{Germany}
}
%

%
%
\begin{abstract}
Numerous studies demonstrated that browser fingerprinting is detrimental to users' security and privacy. However, little is known about the effects of browser fingerprinting on Android hybrid apps -- where a stripped-down Chromium browser is integrated into an app. 
These apps expand the attack surface by employing two-way communication between native apps and the web.
This paper studies the impact of browser fingerprinting on these embedded browsers. To this end, we instrument the Android framework to record and extract information leveraged for fingerprinting. We study over 20,000 apps, including the most popular apps from the Google play store. We exemplify security flaws and severe information leaks in popular apps like Instagram. Our study reveals that fingerprints in hybrid apps potentially contain account-specific and device-specific information that identifies users across multiple devices uniquely. Besides, our results show that the hybrid app browser does not always adhere to standard browser-specific privacy policies.

	\end{abstract}

\maketitle              

\section{Introduction}
Browser fingerprinting is an effective method to identify individuals based on information accessible through browser settings without storing information locally, e.g., in cookies. Web pages capture information about the user and the environment, such as the timezone, locale, and other distinguishable information. Several websites leverage browser fingerprinting to detect botnets and other harmful activity, such as an account accessed from a different place or device than usual. On the flip side, online entities exploit fingerprinting to develop targeted advertisements, price inflation for identified individuals, and targeted malware for particular browser/operating system versions. 

Multiple studies~\cite{Panoptclick,browserfingerpintLargeScale,CanvasFingerprintngAcar,Cao2017CrossBrowserFV, Mowery2012PixelP,AudioEngelhardt,BrowserExtensions, BrowserExtensions2, BrowserExtensions3} acknowledged the privacy and security implication of this topic in the last decade. The majority of these studies targeted desktop browsers, however, recent years have seen a technological shift towards mobile devices rather than desktop PCs for internet browsing. 
A recent study~\cite{oliver2018fingerprinting} explored fingerprinting on mobile browsers and demonstrated fingerprinting to be quite effective on mobile browsers. However, to the best of our knowledge, there are no studies to understand the impact of fingerprinting on hybrid apps.

Hybrid mobile apps integrate native and web components into a single mobile application. Hybrid apps, on the surface, are native applications combined with web technologies such as JavaScript. Hybrid apps offer advantages to developers as they facilitate reusability across multiple platforms: Existing web apps, e.g., login pages, may effortlessly be integrated into multiple mobile platforms (e.g., iOS and Android) to save time and development costs. In this work, we explore the implications of browser fingerprinting on Android hybrid apps. Android framework provides the WebView~\cite{webview} class to integrate hybrid apps. WebView embeds web applications into a view of the Android app and displays webpages in a Chromeless browser~\cite{dumm}.

WebView also provides an active communication channel between the native Android app component and JavaScript in the browser. JavaScript can access the Android app's functionality through shared objects. This grants web components strong capabilities of accessing native Android APIs without having to ask for the Android permissions individually. In contrast to Android's permission system, where users can authorize permissions just once (perhaps in a completely different context), on the web, users must approve sensitive access (e.g., location access) or grant it for one day. However, a hybrid app's inbuilt browser inherits this permission (if the shared Android component has this permission) without further user interference. There have been multiple studies~\cite{rizzo2018babelview,AdLib,LuDroid-Journal,Mutchler2015ALS} to understand the security and privacy implications of Hybrid apps in Android. These studies demonstrated multiple scenarios where hybrid apps are insecure with respect to users' security and privacy. Many hybrid apps use insecured protocols and send private information to third-parties. Unfortunately, the impact of fingerprinting the hybrid app's inbuilt browser is still unknown. 

In this work, we bridge the gap in understanding the impact of hybrid apps’ browser fingerprinting. We perform a large-scale study of fingerprints generated by hybrid Android apps. In particular, we are interested in information leakage, user tracking, and security implications arising from the bridge communication capabilities of hybrid apps. The bridge communication provides (potentially untrusted) web components of hybrid apps access to the trusted native app’s data and functionality. In this work, we explore how the web counterparts of a hybrid app exploit these capabilities to expose information via fingerprinting. Besides, we identify the differences in fingerprinting between the stand-alone and the browser in hybrid apps. To this end, we study over 20,000 apps, including the most popular apps from the Google play store. To obtain the fingerprint of the hybrid app’s browser, we employ dynamic instrumentation of WebView using the Frida instrumentation framework~\cite{frida}. Frida provides a dynamic instrumentation toolkit to inject code into the Android Framework programmatically. In particular, Frida supports overloading of existing methods of the Android Framework. We develop a tool, \toolname, based on Frida to identify and collect the browser fingerprints. \toolname~instruments the Android framework to overload the \emph{loadUrl}, \emph{postUrl} methods of the WebView class, and the \emph{onLoadResource} method of \emph{WebViewClient}. In particular, the instrumentation is targeted to collect three key pieces of information; User Agent String, custom headers, and URLs. URLs help identify the unencrypted traffic originating from \emph{loadUrl}. Custom headers and the User Agent String help identify privacy leaks and unique identifiers associated with the web request. Finally, we exemplify the security flaws and information leaks on popular apps like Instagram. In summary, our study reveals that some apps' fingerprints contain account-specific and device-specific information that can be used to identify and link their users over multiple devices uniquely. Besides, our results show that the hybrid app browser does not always adhere to standard browser-specific privacy policies.

To summarize, this study contributes the following:
\begin{itemize}
	\item{\emph{A Large-scale analysis of Hybrid app’s browser fingerprinting}} We perform a large-scale analysis of the Hybrid app's browser fingerprinting. Our analysis helps to understand the privacy and security implications of fingerprinting on Android hybrid apps. We explore that the hybrid app browser does not adhere to standard browser-specific privacy policies due to customization inability. Besides, many popular apps' fingerprints contain account-specific and device-specific information that can be used to identify their users over multiple devices uniquely. 
	\item{\emph{\toolname}} We develop a tool, \toolname, based on Frida to identify and collect the browser fingerprints. We make our tool public~\cite{zenodo} for the researchers to reuse and build upon it. 
	\item{\emph{Dataset}} We open-source all the datasets~\cite{zenodo}  used in our study to help the researchers and developers to reproduce and understand the implication of fingerprinting on hybrid Android apps.
\end{itemize}

\section{Motivation and Background}
Before delving into the details of our core framework and the implications of browser fingerprinting in Android hybrid apps, we provide a brief background of the techniques utilized in our study.
\subsection{Hybrid Apps}
Android hybrid applications embody native Android parts along with web components. These apps enable developers to reuse their existing web applications in their Android apps. To enable hybrid apps, Android provides a set of APIs to facilitate the communication among Android native app components (primarily written in Java or Kotlin) and web components. These APIs are composed via the Android \emph{WebView} class, which allows the developer to display web pages as a part of the app’s activity (e.g., login screen).

\begin{lstlisting}[caption={Android Hybrid app communication}, label={lst:setter-method-called-from-an-app}, language={JavaScript}, float]
(* \rule[0cm]{8cm}{0.4pt}*)
//Android side: exposing functionality to JavaScript
(*\label{Alstlisting:1}*)public class BridgedClass {
(*\label{Alstlisting:2}*)	public String name;
	
(*\label{Alstlisting:3}*)	@JavascriptInterface
(*\label{Alstlisting:4}*)    public void setValue(String x) {
(*\label{Alstlisting:5}*)    	this.name = x;
(*\label{Alstlisting:6}*)    }
    
(*\label{Alstlisting:7}*)    public String getValue(){
(*\label{Alstlisting:8}*)    	 return this.name;
    }
}
//Activity implementing WebView
@Override
(*\label{Alstlisting:9}*)protected void onCreate(Bundle savedInstanceState) {
//some  code
(*\label{lstlisting:10}*)	WebView wv = (WebView) findViewById(R.id.webview);
(*\label{lstlisting:11}*)	WebSettings webSettings = wv.getSettings().setUserAgentString("My User agent");
(*\label{lstlisting:12}*)	webSettings.setJavaScriptEnabled(true);
(*\label{lstlisting:13}*)	BridgedClass bClass = new BridgedClass();
//share the bridge object to JavaScript
(*\label{lstlisting:14}*)	wv.addJavascriptInterface(bClass, "sharedJavaObject");
// JavaScript invoking Android via the shared object
(*\label{lstlisting:15}*)	wv.loadUrl("javascript:" + "sharedJavaObject.setValue(\"Hello World\")");
// Invoking JavaScript methods
(*\label{lstlisting:61}*)  wv.loadUrl("javascript:set()");
//Loading a url
(*\label{lstlisting:16}*)	wv.loadUrl("http://www.dummy.com");
(* \rule[0cm]{8cm}{0.4pt}*)
//JavaScript side
set() {
(*\label{lstlisting:17}*)  x = new Object(); 
(*\label{lstlisting:18}*)  const str = new String();
(*\label{lstlisting:19}*)  x.f = str.concat("x", "y"); 
(*\label{lstlisting:20}*)  v = x.f;
(*\label{lstlisting:21}*)  sharedJavaObject.setValue(v)
}

\end{lstlisting}
    
WebView provides two styles of communication channels between Android and the web. In the first type, an app can invoke a webpage/script without sharing any Android functionality with them. In the second, more interesting \emph{two-way} communication channel, an app actively communicates with a webpage/script by sharing Android-side functionality to the WebView. The example in \Cref{lst:setter-method-called-from-an-app} contains both of these cases. \Cref{lstlisting:13} and \Cref{lstlisting:14} present the code (using the \lstinline|addJavascriptInterface| API) to share an Android object to JavaScript. In our example, \Cref{Alstlisting:1} to \Cref{Alstlisting:8} describe a class \lstinline|BridgeClass| shared with JavaScript. By default, none of the methods in a class are exposed to JavaScript. The Android framework provides the \lstinline|@JavascriptInterface| annotation to specify the shared methods of a bridge class. For example, \lstinline|BridgeClass| does \emph{not} share the \lstinline|getValue| method to JavaScript. 
\Cref{Alstlisting:9} to \Cref{lstlisting:16} present an Android activity code that creates a WebView. \Cref{lstlisting:10} and \Cref{lstlisting:11} provide a general configuration for creating a WebView. By default, the execution of JavaScript is disabled in a WebView. Developers need to manually enable JavaScript by utilizing \lstinline|setJavaScriptEnabled(true)| (e.g., \Cref{lstlisting:12}). Once enabled, the JavaScript can be invoked using the \lstinline|loadUrl| method. \Cref{lstlisting:15} to \Cref{lstlisting:61} describe two ways to achieve this. Finally, \lstinline|loadUrl| can also be used to invoke normal URLs, e.g., \Cref{lstlisting:16}. 

\subsubsection*{WebView APIs}
WebView provides the following APIs to fetch URLs and execute \emph{JavaScript} scripts.
\begin{itemize}
    \item \emph{loadUrl(Url)}: It loads the specified Url in the WebView. \emph{loadUrl} can also execute JavaScript code. JavaScript script strings are prepended with \texttt{javascript:}.
    \item \emph{loadUrl(Url, HttpHeaders)}: It has the same functionality as \emph{loadUrl} with additional HTTP headers. Developers can specify the HTTP headers they want to bundle with the request. 
    \item \emph{postUrl(Url, postData)}: It loads the specified network \emph{Url} using the POST method along with the post data.
    \item \emph{WebViewClient.onLoadResource(webView, Url)} It notifies the host application that WebView \emph{webView} will load the specified \emph{Url}.
\end{itemize}

\subsubsection*{WebView User Agent Settings}
WebView provides an API to set custom user-agent settings for the WebView browser. Developers can override the user-agent settings, which can be intercepted by the loaded URL. For example, \Cref{lst:setter-method-called-from-an-app} sets the user agent settings to ``My User agent'' (\Cref{lstlisting:11}). 

User-agent settings are useful for user's security, as well as notorious for breaking it. However, the user agent settings in WebView are a bit different from those on browsers. Recently, desktop and mobile browsers, such as Chrome, Mozilla, and others, allow users to hide sensitive information to evade fingerprinting. However, this provision is lacking in the case of WebView browsers. Here, the control is directly in the hands of the developer. This makes WebView browsers a lucrative option for fingerprinting since these may inherit privacy-sensitive data with the shared native Android app’s functionality. Our study shows that developers have leveraged these features to collect users' device fingerprints.

\subsection{Browser Fingerprinting}

Browser fingerprinting is a technique to profile users to uniquely identify them based on passive information, known as a \emph{browser fingerprint}, obtained from the browser. Browser fingerprint uses the information collected from browsers, such as HTTP headers (such as User Agents and Accept), Flash plugins, JavaScript cookies, and many others. Recent advances in the web, such as browser extensions, canvas elements, and WebGL components are also known to be sources of fingerprints~\cite{bsurvey,CanvasFingerprintngAcar}. We explain three approaches here: (1) User Agents, (2) Accept and Content-Language, and (3) browser extensions to aid the understanding of this paper for our readers. Interested readers may refer to  Laperdrix~\emph{et al.}~\cite{bsurvey} for a detailed survey of browser fingerprinting.


The HTTP protocol is meant to be platform-independent, and therefore, browsers rely on the information from HTTP headers to identify the browser of an incoming request. The information is encoded in the standard HTTP semantics (RFC 9110~\cite{rfc9110}) called as \emph{User-Agent request headers} or User Agent strings. User-Agent strings specify the system characteristics such as browser, operating system, architecture, and many others, and are used by web servers to identify the client information. As of now, User-Agent strings are complex and add a plethora of information other than the browser. Developers can override the existing user-agent headers and inject information into these headers. For example, JavaScript facilitates developers to modify these strings and add more information, such as timezone, screen-specific attributes (such as resolution, depth), platform, and many others. This information is a source of fingerprints as shown by earlier works~\cite{bsurvey, Panoptclick}. 

Accept headers are used to specify the file types accepted by the browsers is another source of fingerprintg~\cite{bsurvey, Panoptclick}. It is a comma-separated list of content types and their subtypes. For example, a browser can set the accept headers to \verb|text/html, application/xhtml+xml|, which indicates the browser can accept the type \texttt{text} of sub-type \texttt{html}. Content-Language attribute specifies the localization information of the browsers, such as \texttt{de-DE, en-US, en-IN}. Content-language is also a source of localization information for fingerprinting~\cite{AmIUnique}.

Browser extensions are browser-based applications that enhance the browsing experience. Although these improve browser experience, such as by reducing ads, they are also a source of fingerprinting information. Starov and Nikiforakis~\cite{BrowserExtensions} identified 14.10\% of users via fingerprints obtained from their browser extensions. They used the changes in the DOM model introduced by the browsers to detect extensions. A similar study from Sanchez-Rola et~al. showed the possibility of extension enumeration attack on browsers, thus identifying 56.28\% users from 204 users. To this end, they measure the timing difference between querying resources of fake and benign extensions.




\paragraph{Large  scale studies  on browser fingerprinting}
Browser fingerprints can compromise users' privacy. It was first demonstrated in the experiment \emph{Panoptclick}~\cite{Panoptclick} by Peter Eckersley from the Electronic Frontier Foundation, where he fetched around 470,000 fingerprints, of which around 84\% were unique. His experiment shows the gravity of the problem, i.e., browser fingerprints can uniquely determine a majority subset of the users on the web. Following up on these experiments, researchers revealed many other sources of browser fingerprinting generation techniques to profile users and break their privacy. We list these techniques in the related work of this paper.

The evolution of the Web from desktop to mobile browsers has affected users' privacy from browser fingerprinting. Earlier research~\cite{oliver2018fingerprinting} shows that fingerprints from mobile browsers reveal a lot more sensitive information than from desktop browsers. To tackle the problem of fingerprinting, web browsers have started introducing policies to minimize browser fingerprints. Unfortunately, these policies do not apply to the hybrid app's in-built browser, leaving the control in the hands of the developers. We study this aspect in this paper.

\paragraph{Uniqueness of a fingerprint.} To compare the strength of the information revealed by the fingerprints obtained in our study, we compare it against a larger dataset of the \emph{Cover your tracks}. It shows the bits of unique information revealed by the fingerprint, which matches with the fingerprint obtained in our database. \emph{Cover your tracks} shows this information in terms of the number of browsers having the same fingerprint. In this paper, we refer to it as \emph{uniqueness}. 



\section{Methodology}
\begin{figure}[tb]
	\centering
	\includegraphics[width=\columnwidth]{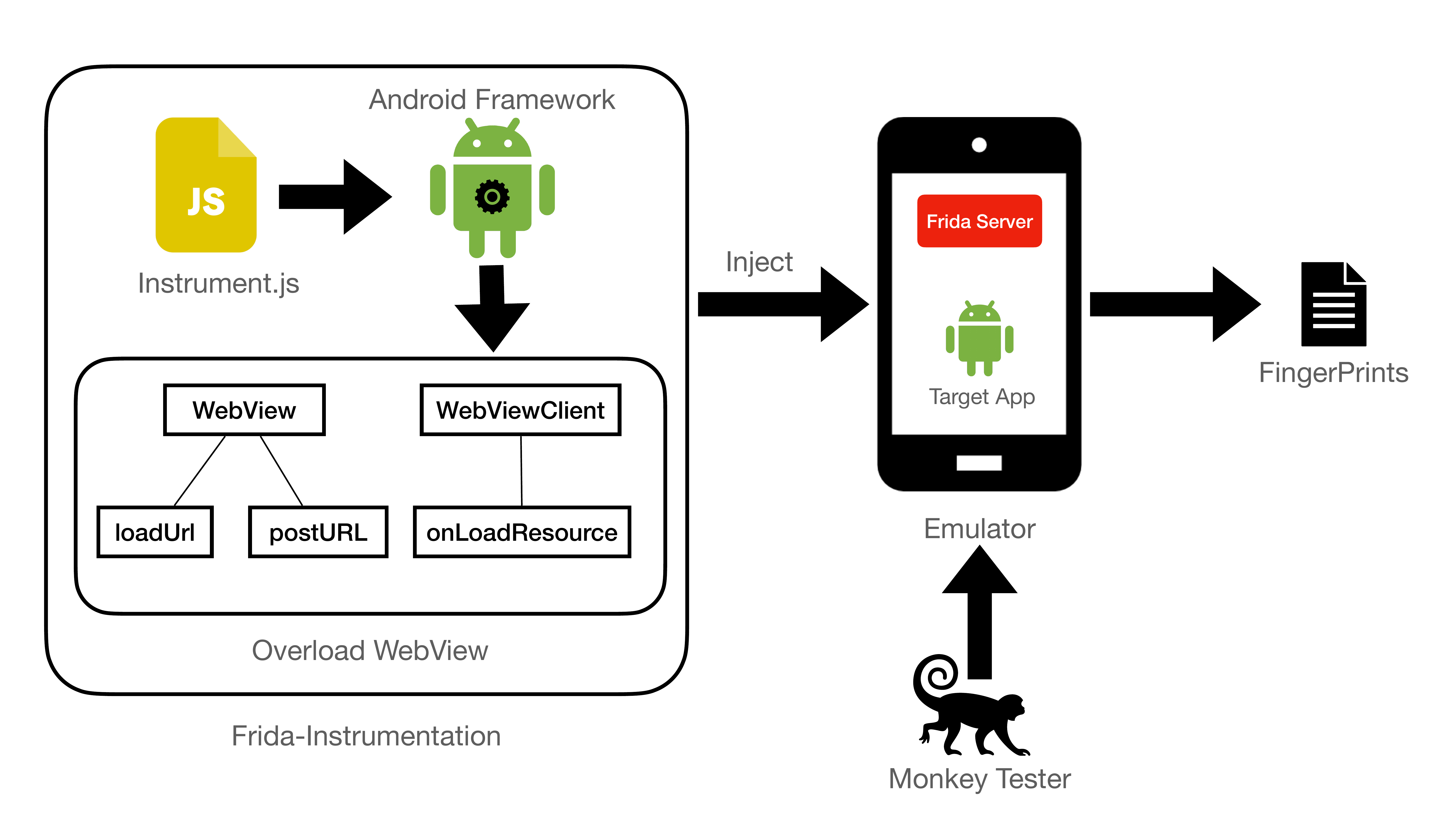}
	\caption{Workflow of \toolname}
	\label{Fig:workflow}
\end{figure}

Unlike fingerprinting in traditional browsers, fingerprinting hybrid apps has inherent technical challenges. With traditional browsers, it is feasible to attach scripts/plugins to a web page and rely on cookies to gather information, which is, unfortunately, not possible with hybrid apps. The hybrid app browser is provided as a part of the Android Framework, and it displays web pages as a part of the app’s activity.
In this work, we perform runtime instrumentation of the \emph{WebView} class to intercept the fingerprinting data. Generally, network analysis tools such as \emph{Wireshark} could also obtain parts of the required data. However, for a large scale analysis, instrumenting the \emph{WebView} class gives us more control over the data we collect, e.g., \emph{Wireshark} does not associate the apps' identifier to the network traffic containing fingerprinting data. Besides, instrumenting  \emph{WebView} enables us to capture the direct traffic from the particular app, while  \emph{Wireshark} captures all traffic, including noise from other apps and Android Framework. 


\autoref{Fig:workflow} provides an overview of our instrumentation framework. There are two potential ways to instrument WebView. First, modifying the Android framework by integrating the required code changes directly into the Android Open Source project, and then running the apps on this custom Android OS. Second, achieving the desired modifications with the help of dynamic instrumentation. In this work, we opt for the latter path; we leverage an existing Android dynamic instrumentation framework, Frida~\cite{frida}. Frida provides a dynamic instrumentation toolkit to inject code into the Android Framework programmatically. In particular, Frida supports overloading the existing methods of the Android Framework. We develop a tool, \toolname, based on Frida to identify and collect the browser fingerprints. \toolname~instruments the Android framework to overload the \emph{loadUr}, \emph{postUrl} methods of the \emph{WebView} class, and \emph{onLoadResource} of \emph{WebViewClient}. In particular, the instrumentation is targeted to collect three key pieces of information; User Agent String, custom headers, and URLs. URLs help to identify the unencrypted traffic originated from \emph{loadUrl}. Custom headers and the User Agent String help to identify privacy leaks and unique identifiers associated with the transmission.
To navigate through various Android activities, we leverage the Android automated tester \emph{Monkey}~\cite{monkey}. Monkey can produce pseudo-random streams of user events such as mouse movements and gestures and generate various system-level events to help in the automatic navigation of the Android apps. 

\autoref{lst:intrumentloadurl} presents the pseudocode for instrumenting the \emph{loadUrl} method with a single parameter. Line~\autoref{lstlisting:1} creates an object \emph{WebView} pointing to the Android Framework's \emph{WebView} class. In the next line, the \emph{loadUrl} method is overloaded to extract browser fingerprints. In Line~\autoref{lstlisting:6}, the app’s unique identifier (package name) is extracted to associate it with the fingerprints. Finally, custom headers, the user-agent string, and URL are extracted and logged in Line~ \autoref{lstlisting:7} and \autoref{lstlisting:8}.

\begin{lstlisting}[caption={Instrumenting the WebView-- Overloading the loadUrl(Url)}, label={lst:intrumentloadurl}, language={JavaScript}, float]
(*\label{lstlisting:1}*)var WebView = Java.use("android.webkit.WebView");
(*\label{lstlisting:2}*)WebView.loadUrl.overload('java.lang.String').implementation = function(url) {
(*\label{lstlisting:3}*)	this.loadUrl(url);
(*\label{lstlisting:4}*)	const ActivityThread = Java.use('android.app.ActivityThread');
(*\label{lstlisting:5}*)	var context = ActivityThread.currentApplication(). getApplicationContext();
(*\label{lstlisting:6}*)	var packagename = context.getPackageName();
(*\label{lstlisting:7}*)	send({ (*\Suppressnumber|*)
		packageName: packagename, 
		method: "loadUrl",
		Url: url,
		Header: "",
		userAgent: this.getSettings().getUserAgentString()
	});  (*\Reactivatenumber*)
(*\label{lstlisting:8}*)	console.log("WebView.loadUrl url:" + url);
}
\end{lstlisting}
\label{sections/methodology}

\section{Evaluation}
\begin{figure}[tb]
	\includegraphics[width=\columnwidth]{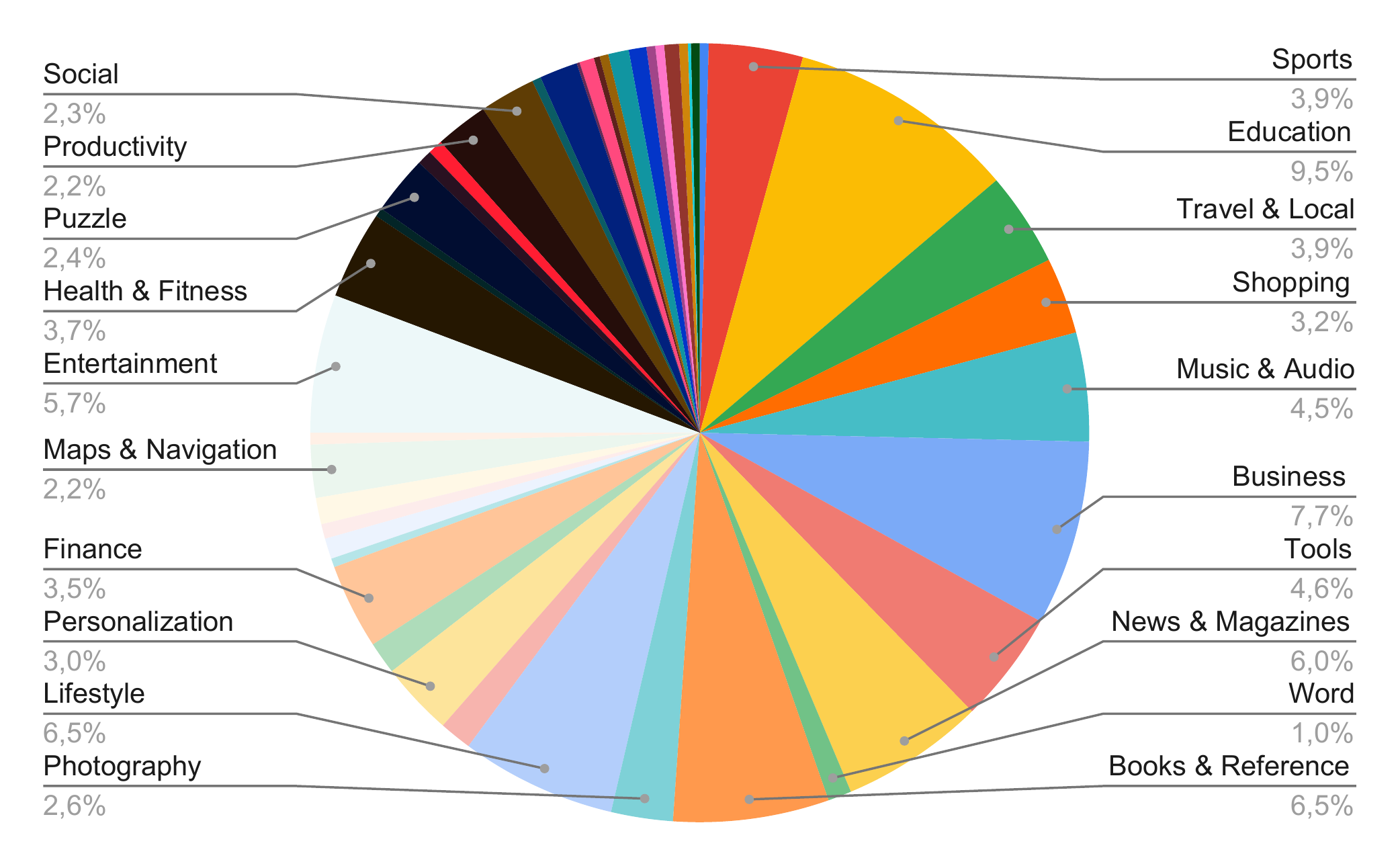}
	\caption{Apps by Categories}
	\label{category}
\end{figure}

\paragraph{\textbf{Dataset}} We conducted our study on over 20,000 apps from the AndroZoo~\cite{allix2016androzoo} dataset. AndroZoo contains a compilation of Android apps from several sources, including the Google Play store. In our study we are interested in hybrid apps, which contain at least one instance of WebView. Thus, to filter for hybrid apps, we first decompiled the apps in the dataset and examined the decompiled code for WebView-related method signatures. To further validate that these apps are hybrid, we applied our instrumentation framework to them and logged WebView-related method calls. We ended up with 5,145 apps that use at least one instance of WebView's APIs. We were also interested in the app store categories of these apps, so we created a script that automatically determines the category of an app in the Google play store based on its package name. This categorization was successful for approximately 1000 apps, the remaining apps are not/no longer listed in the Google play store, which precludes automated classification. Thus, the pie chart in \Cref{category} provides the distribution of categories for the more than 1000 apps (still) available in the Google play store only. 

On top of this dataset, we selected the ten popular apps from the Google Play store (as of April 2022) for automatic as well as manual analysis. In particular, we created multiple (fake) accounts and observed http headers like cookies, user-agent strings, and URLs for these accounts. The manual analysis aims to determine information that can help identify a user uniquely over multiple devices or platforms. \Cref{table:manually-analyzed-apps} lists these ten apps, along with the sensitive information they expose in their user agent, cookies, and custom headers.
 
All of these applications were subsequently instrumented as described in \Cref{sections/methodology} to collect the user agent strings, custom headers, and URLs. We further created scripts to automate the data collection process: All of our scripts are publicly available to researchers for replication purposes.
Our experiments were performed on a personal laptop with 16~GB RAM and a fourth-gen Intel Core i7-4500U processor running Windows~10. 
\begin{table*}[tb]
	\caption{Manually Analyzed Apps}
	\label{table:manually-analyzed-apps}
		\begin{tabular}{l|l|l|l|l|l}
			\hline
			\textbf{App Name} &
			\textbf{Version} &
			\textbf{Category} &
			\textbf{Cookie} &
			\textbf{User agent} &
			\textbf{Custom headers} \\ \hline \hline
			Instagram &    229.0.0.17.118   & Social                   & no  &\begin{tabular}[c]{@{}l@{}} Phone model, build number, \\ localization info, SDK, \\ Android version, processor \end{tabular}  & no        \\ \hline
			Facebook &   359.0.0.30.118     & Social                   & no  & Phone model, build number & no        \\ \hline
			Alibaba   &   7.48.1    & Shopping                 & yes & Phone model, build number & unique ID \\ \hline
			Twitter    &   9.31.1   & Social                   & no  & Phone model, build number & no        \\ \hline
			LinkedIn   &  4.1.629.1   & Social                   & no  & Phone model, build number & no        \\ \hline
			UBer        & 4.361.10001   & Cab                      & no  & Phone model, build number & no        \\ \hline
			QuuBe - Wholesale & 6.5.1  &
			Shopping &
			yes &
			\begin{tabular}[c]{@{}l@{}}Phone model, build number, \\ UUID in the user agent\end{tabular} &
			UUID \\ \hline
			flipboard    & 4.2.97  & Shopping                 & no  & Phone model, build number & no        \\ \hline
			Youtube    &   17.08.32  & Video Players \& Editors & no  & Phone model, build number & no        \\ \hline
			DW Learn German & 1.0.1 & Education                & no  & Phone model, build number & no        \\ \hline
		\end{tabular}%
\end{table*}
\paragraph{\textbf{Case Studies}}
Multiple studies have been proposed for browser fingerprinting~\cite{oliver2018fingerprinting,Panoptclick,bsurvey,browserfingerpintLargeScale} and Android hybrid app analysis~\cite{LuDroid-Journal,rizzo2018babelview,AdLib,Mutchler2015ALS}. The most relevant recent work~\cite{oliver2018fingerprinting} performed a preliminary investigation on fingerprinting of mobile browsers. However, their work focused on full-fledged mobile browsers. In contrast, we aim to perform a large-scale study of fingerprints generated by hybrid Android apps. In particular, we are interested in information leakage, user tracking, and security implications arising from the bridge communication capabilities of hybrid apps. The bridge communication provides access from (potentially untrusted web components of a hybrid app to the trusted native app’s data and functionality. In this work, we explore how the web component of a hybrid app exploits these capabilities to expose information via fingerprinting. Besides, we identify the differences in fingerprinting between the stand-alone and the hybrid apps’ browser. In summary, we find that hybrid apps reveal more information about the user than traditional browsers. We exemplify the research findings in the form of the following case studies:
\paragraph{\textbf{Case Study 1: Privacy leakage unique to hybrid apps' browser.}}

Fingerprints in WebView are a good source of (potentially) privacy-sensitive information. For example, the hybrid app browser’s fingerprint contains sensitive information such as the phone model and build number. The latter is sensitive information that can be leveraged to determine vulnerable devices and craft operating-system-specific attacks as observed by security analysts~\cite{nightwatch} and acknowledged by Google~\cite{chromiumBug}. The desktop Chrome browser removed the build number in 2018 whereas the hybrid apps’ browser includes this information in the user agent string up to this date. 

To further improve user privacy, Chrome contains a privacy sandbox since version 93 (released on August 31, 2021). It allows the user to manually limit\footnote{Via \url{chrome://flags/\#reduce-user-agent}} leaking of sensitive information to protect against passive fingerprinting. 
However, no such configuration can be activated in hybrid apps’ in-built browser. \Cref{fig:chromium-fingerprints} shows the uniqueness of the fingerprints obtained on hybrid apps’ in-built browser, the standalone Chrome browser, and the Chrome browser with sandboxing.  The \emph{uniqueness} brought by the privacy sandbox is 259 times lower than the unmasked fingerprint: The higher the uniqueness number, the worse it is for users’ privacy.
\begin{table*}[bt]
	\caption{Fingerprints from Various Browsers}
	\label{fig:chromium-fingerprints}
		\begin{tabular}{l|@{}|p{0.5\linewidth}|c}
			\hline
			\textbf{Platform} &
			\textbf{Fingerprint} &
			\textbf{Uniqueness (1/X)} \\ \hline  \hline
			Hybrid apps' Browser &
			\texttt\{Mozilla/5.0 (Linux; Android 9; SM-A505FN Build/PPR1.180610.011; wv) AppleWebKit/537.36 (KHTML, like Gecko) Version/4.0 Chrome/99.0.4844.88 Mobile Safari/537.36\} &
			X= 218256 \\ \hline
			Chrome Browser &
			\texttt\{Mozilla/5.0 (Linux; Android 9; SM-A505FN) AppleWebKit/537.36 (KHTML, like Gecko) Chrome/98.0.4758.87 Mobile Safari/537.36\} &
			X = 218112 \\ \hline
			Chrome Browser with sandboxing &
			\texttt\{Mozilla/5.0 (Linux; Android 10; K) AppleWebKit/537.36 (KHTML, like Gecko) Chrome/93.0.0.0 Mobile Safari/537.36\} &
			X=838.98 \\ \hline
		\end{tabular}%
\end{table*}

To obtain the \emph{uniqueness} of a browser fingerprint, we leverage \emph{Cover Your TRACKS}~\cite{tracks}, a research project to understand the uniqueness of browser fingerprints. It provides a uniqueness score to a fingerprint based on a large fingerprint database. We observed that fingerprints including the build number are highly unique; the uniqueness decreases significantly when removing the build number, and again drastically when limiting the phone model information.

\vspace{5pt}
\noindent\framebox{\parbox{\dimexpr\linewidth-2\fboxsep-2\fboxrule}{%
	\textbf{Finding 1:}	\itshape \, Hybrid apps’ built-in browser permits more sensitive information leakage than the stand-alone browser.  All hybrid apps in our dataset expose the build number and phone model in their fingerprints. This permissiveness stems from the inability to configure system-wide privacy policies.}}
\vspace{3pt}

\paragraph*{\textbf{Case study 2: Information leak by Instagram app.}}

Like traditional browsers, Android allows WebView to transmit a user-agent HTTP header to the server, which can derive information from it. It is the app developers' responsibility to control the information they want to share with the server. As is, the web components (WebView) of hybrid apps indirectly inherit the same level of permissions as the shared components of the native side of the apps. Thus, by using the shared APIs, they potentially have access to  sensitive device/user-specific information. During our manual analysis of the most popular apps from the Google play store, we observed an interesting mechanism to profile users based on the HTTP headers in the well-known social media app Instagram. Instagram’s Android app leverages WebView to open an in-app URL/link, i.e., a link shared in a chat. We crafted a scenario where a curious (or malicious) user, Bob, wants to get some personal information such as the phone model, language, or ethnicity of a user Alice. Bob owns a server that can create account-specific links (e.g., server.com/Alice) and sends this link to Alice, and once Alice clicks on this link, it is displayed in the built-in WebView browser. \Cref{fig:fingerprint-instagram} shows the fingerprint and the sensitive information shared with Bob’s server; Bob is able to obtain Alice’s personal information, such as phone model and language preferences. In particular, this attack is plausible in any app that uses WebView to open in-app URLs.

\begin{figure*}[tb!]
\centering
\begin{subfigure}{\linewidth}
\begin{LVerbatim}[frame=single]
Mozilla/5.0 (Linux; Android 9; SM-A505FN Build/PPR1.180610.011; wv) AppleWebKit/537.36(KHTML, like Gecko) 
Version/4.0 Chrome/99.0.4844.88 Mobile Safari/537.36 Instagram 229.0.0.17.118 
Android (28/9; 420dpi; 1080x2131; samsung; SM-A505FN; a50; exynos9610;en_DE; 360889116)
\end{LVerbatim}
\caption{Fingerprints}
\end{subfigure}
\begin{subfigure}{\linewidth}
	\begin{LVerbatim}[frame=single]
Instagram Version (Instagram 229.0.0.17.118, 360889116), Platform (Android)
Android SDK (28) and version (9), Phone model (samsung; SM-A505FN;),  Proccessor name (exynos9610)
DPI and Resolution (420dpi; 1080x2131), Locale (en_DE)
	\end{LVerbatim}
\caption{Identifying Information}
\end{subfigure}
\caption{Fingerprint from Instagram}
\label{fig:fingerprint-instagram}
\end{figure*}
As discussed in case study 1, the Instagram app, by default, sends the phone's model and build number, already providing more uniquely identifiable information than the stand-alone Chrome browser. On top of that, it also reveals the Android version (both OS and SDK), phone resolution, processor name, and localization information. Localization information is very sensitive for profiling users. We observed that the uniqueness of this information is very high (217923), which is detrimental to users' privacy.

This fine-grained information in the user-agent header renders the app vulnerable to passive fingerprinting, where an attacker can infer these user-agent headers by simply observing the traffic coming from a malicious URL shared through the chat. To mitigate the problem of passive fingerprinting, RFC9110~\cite[ch. 10.1.5]{rfc9110} 
disallows ``generate advertising or other nonessential information within the product identifier''. Instagram adds personally identifiable information to the contrary. In contrast to the stand-alone browser where the user can choose to hide this information, the user has no control over which information is shared once certain permissions are given to the Instagram app. 

\vspace{5pt}
\noindent\framebox{\parbox{\dimexpr\linewidth-2\fboxsep-2\fboxrule}{%
		\textbf{Finding 2:}	\itshape \, Hybrid apps are susceptible to passive fingerprinting and often violate standard privacy policies. Famous apps like Instagram provide less to no control to their users over the amount of sensitive information released via web components.}}
\vspace{3pt}

\paragraph{\textbf{Case study 3: Profiling Users via a combination of cookies and user-agent.} } In the previous case studies, we demonstrated how users could be profiled based on user-agent strings. The situation becomes more severe when this information is combined with other mediums such as cookies; the combined information helps obtain a fine-grained profile of the user.  For example, in the Alibaba app, the user's account ID (unique over multiple devices) is added to the cookies; thus, one can intercept the user ID and the phone model information obtained from the user-agent string to profile users' phone buying behavior. Note that the user's account ID stays the same over various devices/browsers, i.e., users can be uniquely identified over different service providers. Besides, the server can concretely infer sensitive information on the user, e.g., how many devices a user owns, how frequently users change their phone, and what the financial situation of a user is. 

User profiling is also possible through HTTP ACCEPT-language headers. ACCEPT-language headers are used to determine the language preferences of the client. Generally, these headers are derived from the language preference of the user. For example, a user located in Switzerland and speaking German would have the accept language \emph{CH-de}. Unfortunately, a user can be profiled based on her language preferences, e.g., identifying the user’s origin, ethnicity, or nationality. Worse, if the user speaks more languages, with the combination of other fingerprintable information, the user can be uniquely identified.  For example, a user speaking a combination of Russian and Turkmen languages could be profiled as Turkmenistan origin. However, users can hide this information on regular browsers through their settings or, better, use a privacy-compliant browser. Unfortunately, this is not possible for the hybrid browser as users cannot control the settings of this browser.

\begin{table*}[t]
		\caption{Apps including unique IDs into user-agent string}
	\label{tab:uniqueId10}
		\begin{tabular}{l|l|l}
			\hline 
			\textbf{Package Name}                  & \textbf{App Name}             & \textbf{Category} \\ \hline \hline
			com.oddm.adpick                        & Adpick                        & Office            \\ \hline
			net.giosis.shopping.id                 & Qoo10 Indonesia              & Shopping          \\ \hline
			net.giosis.shopping.cn.nonepush        & Qoo10  APK 3.2.7         & Shopping          \\ \hline
			Net.giosis.shopping.sg                 & Qoo10 - Online Shopping 6.5.1& Shopping          \\ \hline
			xyz.quube.mobile                       & QuuBe - Wholesale by Qoo10    & Shopping          \\ \hline
			xyz.quube.shopping.tablet              & QuuBe for Tablet              & Shopping          \\ \hline
			mobile.qoo10.qpostpro                  & Qpost Pro 1.4.1               & Shopping          \\ \hline
			mobile.qoo10.qstl20                    & Style Club 6.4.0              & Shopping          \\ \hline
			com.alibaba.intl.android.apps.poseidon & Alibaba                       & Shopping          \\ \hline
			Com.accelainc.ihou.fr.droid            & Illegal Romance 1.0.2         & Adventure         \\ \hline
		\end{tabular}%
\end{table*}
Furthermore, we observed that various applications attach unique device IDs to the user-agent string, resulting in the direct identification of a user. To observe this behavior, we logged into the apps with multiple user accounts and observed the differences in the fingerprints. This manual analysis confirms this misconduct~\cite[ch. 10.1.5]{rfc9110} in at least ten apps in our dataset. \Cref{tab:uniqueId10} presents the list of these apps alongside their categories. Apps with a similar name, e.g., Qoo10 Indonesia and Qoo10 APK 3.2.7, are from the same manufacturer but belong to different countries and have different privacy policies. Owing to the sheer volume of the dataset, it was not feasible to create multiple accounts for all the apps and relate fingerprints for this unique information. \Cref{fig:unique-id-in-user-agent} shows a sample of the fingerprints obtained from the devices containing unique device IDs. As is, the unique IDs are attached to the devices; they remain unchanged after even reinstalling the apps. Along with the unique device ID, these devices contain fine-grained information about the device attributes, such as build number, phone model, and Android version. Thus, one can directly relate a device to its attributes, and also build a temporal profile of the particular device, in case the device is used by another user.
\begin{table*}[tb!]
	 \caption{Fingerprint showing unique ID}
	\label{fig:unique-id-in-user-agent}
    \begin{tabular}{p{0.28\linewidth}|l|p{0.6\linewidth}}
        \hline
        \textbf{App} &  \textbf{Category} & \textbf{Fingerprint} \\ \hline \hline
        com.oddm.adpick & Office & 
\lstinline[breaklines=true, style=fingerprint]!Mozilla/5.0 (Linux; Android 10; Android SDK built for x86 Build/QSR1.210802.001; wv) AppleWebKit/537.36 (KHTML, like Gecko) Version/4.0 Chrome/74.0.3729.185 Mobile Safari/537.36 AdpickEncrypted:GDPViCyiXnbcQgWnvAmIBusjAV43FvgPeawc /Xc5ayQW0rBy/oA8BUz4Vdmy9ITgwRQDnaI7BmZB#nXG5+MzNecK3 HyqXv7P5/2u9yqMmkwrA/leTfsNeUZbmjvzj9D9m ECLyuBwl3lA8Sz 2dt4Ue1H1tT#n4mWgFssSh2n/eR1qpgnGRhc1cB2jqXtWuTW/cNQC#n! \\ \hline

 net.giosis.shopping.id  &  Shopping &
\lstinline[breaklines=true, style=fingerprint]!Mozilla/5.0 (Linux; Android 11; Android SDK built for x86 Build/RSR1.210210.001.A1; wv) AppleWebKit/537.36 (KHTML, like Gecko) Version/4.0 Chrome/83.0.4103.106 Mobile Safari/537.36 Android_Gmarket Qoo10 ID_3.6.2_133(GMKTV2_ZlRnG1XAIzgwoC3OBe0hNjV4PfmyaC5RAIBqY+ mkcipUGsSIiB19AyfIHQY1msEafG/xGz9RIS4=;Android SDK built for x86;11;en_US;2000010476)! \\ \hline

net.giosis.shopping. cn.nonepush & Shopping &
\lstinline[breaklines=true, style=fingerprint]!Mozilla/5.0 (Linux; Android 10; Android SDK built for x86 Build/QSR1.210802.001; wv) AppleWebKit/537.36 (KHTML, like Gecko) Version/4.0 Chrome/74.0.3729.185 Mobile Safari/537.36 Android_Gmarket Qoo10 CN NOPUSH_3.6.6_137(GMKTV2_/E/eowDAPJdLOH3or4b6kUZaqiQ9445kf5 0bcLzkkQeoFvJmnsEzdFnnyGmoyagfCYHYKlwCWP4=;Android SDK built for x86;10;en_US;2000000134)! \\ \hline

net.giosis.shopping.sg & Shopping & \lstinline[breaklines=true, style=fingerprint]!Mozilla/5.0 (Linux; Android 10; Android SDK built for x86 Build/QSR1.210802.001; wv) AppleWebKit/537.36 (KHTML, like Gecko) Version/4.0 Chrome/74.0.3729.185 Mobile Safari/537.36 Android_Gmarket Qoo10 SG_6.5.1_269(GMKTV2_yQ+4mthiJO62KzrgMNh9rwIUgQVt5Aax6jISAX Y3h++KFBJ4DO5 /YZdeiP3jYmD+hnf246qDDdk=;Android SDK built for x86;10;en_US;200007873;US;)! \\ \hline
 
com.accelainc.ihou.fr. droid & Adventure & 
\lstinline[breaklines=true, style=fingerprint]!2NSsaFdT 60D1F74326F469CB__5DC12396C15AB57696B4A 69152169D 1.0.1 & Mozilla/5.0 (Linux; Android 10; Android SDK built for x86 Build/QSR1.210802.001; wv) AppleWebKit/537.36 (KHTML, like Gecko) Version/4.0 Chrome/74.0.3729.185 Mobile Safari/537.36! \\ \hline

    \end{tabular}
\end{table*}
\noindent\framebox{\parbox{\dimexpr\linewidth-2\fboxsep-2\fboxrule}{%
		\textbf{Finding 3:}	\itshape \, The combination of cookies and user agents links sensitive device and user-specific information. This information can be exploited to profile a user uniquely, such as identifying the origin and estimating the personal financial status. Besides, a few apps in our dataset attach their users’ account IDs (unique for a user) to the cookies making their users uniquely identified over different devices.}}
\vspace{3pt}

\begin{figure}[tb]
	\includegraphics[width=\columnwidth]{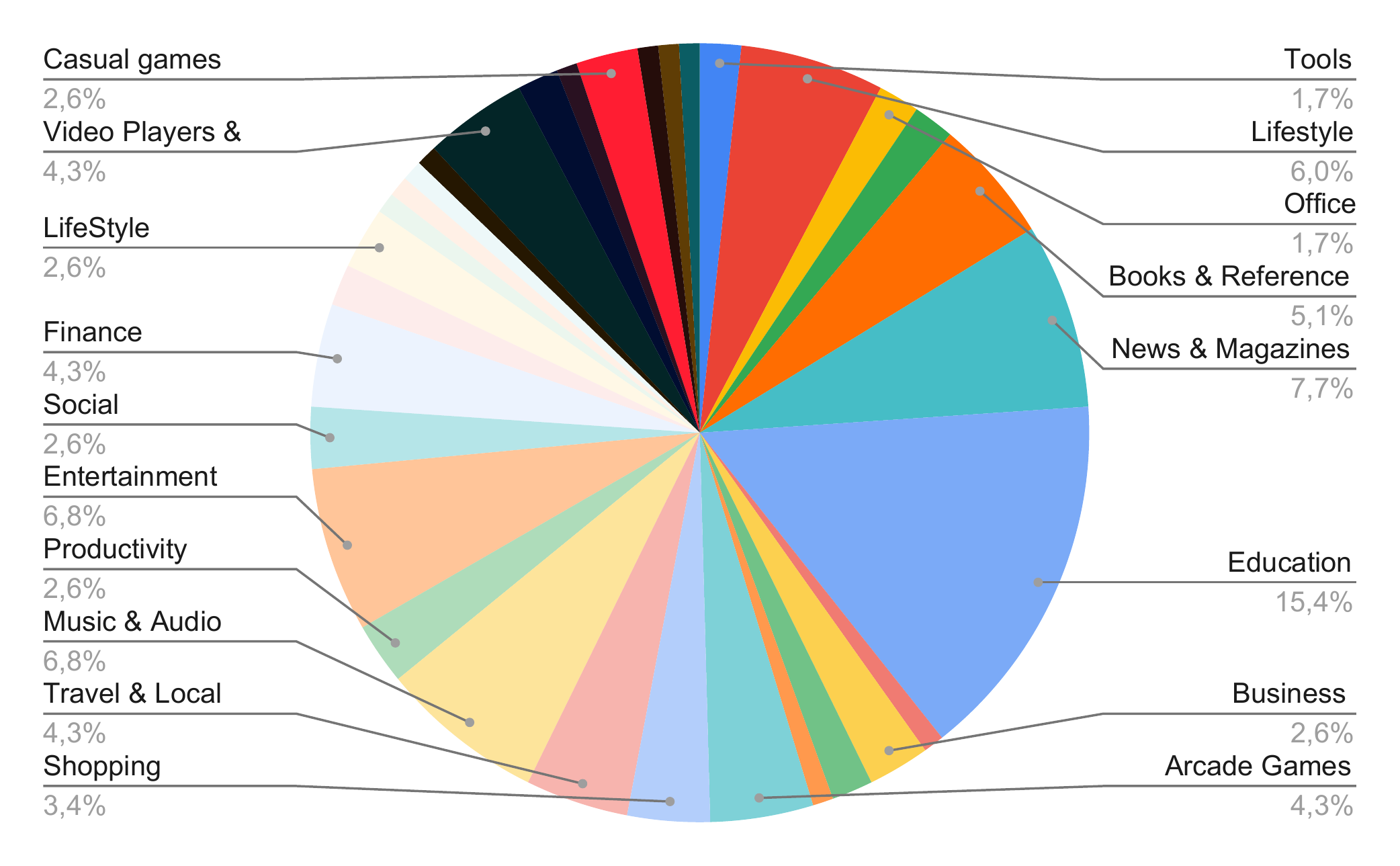}
	\caption{Unencrypted URLs by App Categories}
	\label{unUrl}
\end{figure}

\paragraph{\textbf{Case Study 4: JavaScript modifying Android objects.}}
    \begin{lstlisting}[breaklines=true, ,numbers=none,  basicstyle=\footnotesize\ttfamily, label=fig:device-ids-JavaScript, caption={Setting device IDs through JavaScript}, float]
	JavaScript:if(window.Application) 
	{ 
		Application.setDeviceUid(""APA91bG956w4WPzLIh
		DCHdcnIdbigwApzJzX-WFCkrKRcpJMr9Xw0kbAAxjBYj-
		f6UnVrfeMWRhuPlQIiv8np8733GgHzHm6QHLMeK1
		-InIkhWvxq9yjGb_i2a5WdxIQmaAl-QP3aHHIqK9XTGJiiPpJo
		_dXqkVNzQ"");
	}
\end{lstlisting}
As a part of our instrumentation framework, we instrument the \emph{loadUrl} method to extract the originating URLs. On top of loading URLs \emph{loadUrl} also provides functionality to load/execute a JavaScript code snippet directly. We also intercepted many cases where JavaScript modifies Java objects using bridge objects. A recent study~\cite{LuDroid-Journal} exposed instances of potentially untrusted JavaScript code interfering with Android objects. However, in several cases, the aim of such interference was unknown in that study. In this work, we identify a number of patterns where JavaScript transmits unique IDs to native Android objects. These unique IDs can be used as fingerprints for devices. For example, an app \emph{com.a2stacks.apps.app57191abb7ab09} sets the user ID of the user as shown in \cref{fig:device-ids-JavaScript}, violating multiple security policies. First, the (potentially) unsafe web component violates the integrity of the native app by modifying its object, i.e., writing the device UID into a field. Second, the app may violate the Android privacy policies by assigning a unique device identifier without having asked for permissions. 
\noindent\framebox{\parbox{\dimexpr\linewidth-2\fboxsep-2\fboxrule}{%
		\textbf{Finding 4:}	\itshape \, (Potentially) Unsafe web components infringe the integrity of a native app's object. Hybrid app web components (JavaScript) assign unique identifiers to the device for (potential) fingerprinting purposes via the Android bridge communication. }}
\paragraph{\textbf{Case Study 5: Unencrypted communication.}}
During our analysis of extracted URLs, we find various instances where unencrypted protocols such as HTTP are used to communicate secret information such as device IDs, IP addresses, Google ads user identifiers, and many other sensitive data. This is a severe problem, and unfortunately, 1646 applications from our dataset contain this flaw. Related work~\cite{LuDroid-Journal} has shown that the use of unencrypted communication is susceptible to simple man-in-middle attacks: An attacker can alter the server’s response to an attacker-controlled web page without the user noticing any difference. Besides, the attacker learns the user's sensitive information by just observing the traffic; 281 apps share Google ads IDs, and 132 out of them also add IP addresses to the URLs. Interestingly, 214 of these 281 apps use URLs from the domain \url{http://splash.appsgeyser.com} domain, 28 from \url{http://splash.appioapp.com}, and 39 from \url{http://ads.appioapp.com}.  Note that, all of these URLs belong to platforms (AppsGeyser and Appio) for creating Android apps and the use of unencrypted communication is susceptible to many other apps (not in our dataset). \Cref{fig:unencypted-communication} shows a list of twenty apps that load at least one instance of an unencrypted URL. \Cref{unUrl} provides the distribution of apps\footnote{Over 200 apps that we could categorize.} using unencrypted URLs based on categories. 

\begin{table*}[tb]
	\caption{Twenty Apps with Unencrypted URLs}
	\label{fig:unencypted-communication}
	
	%
		\begin{tabular}{l|l|l|l}
			\hline
			\textbf{Package Name}                  & \textbf{App Name}  &   \textbf{Hash}          & \textbf{Category}  \\ \hline  \hline
			com.wWelcometoPurnia                   & Welcome to Purnia     &   004BDEAF41      & Lifestyle          \\ \hline
			com.wPBALogistics                      & PBALogistics APK     &    0094D388AB      & Office             \\ \hline
			com.wKPUKabKepulauanSelayar            & KPU Kab Kepulauan Selayar &   005F8F4E97  & Communication      \\ \hline
			com.wsmile2                            & Smile APK 1.1      &   00F784BF5B         & Communication      \\ \hline
			com.wAnEssayonManmoralessaysandsatires & An Essay on Man APK       &  0183B4DF5C   & Books \& Reference \\ \hline
			com.cultplaces                         & Cult Places       &     0665508043        & Travel \& Local    \\ \hline
			com.wTrendyBotswana                    & Trendy Botswana       &      067999FD77   & News \& Magazines  \\ \hline
			com.wProfDrMustafaKaratasSoruCevap     & Mustafa Karataş ile Soru Cevap   & 06F06AFCB9  & Lifestyle          \\ \hline
			com.wTanksDecades                      & Tanks Decades       &    06F781FF93       & Arcade Games       \\ \hline
			com.wRapKlayBBJ                        & Rap Klay BB.J         &   0711A4A1AE      & Music \& Audio     \\ \hline
			com.wCaringForYourCat4          & Caring For Your Cat 1.0 & 03BA1A25F1  & Books \& Reference \\ \hline
			com.wMayankCreation          & Mayank Creation 0.1  & 03EB88AF & Lifestyle\\ \hline
			com.wSwiftSpaceship          & Swift Spaceship & 047AE609BB & Arcade \\ \hline
			com.wTamilNaduSSLCResult2016         & Tamil Nadu SSLC Result 2016 0.9 & 04DB332396 & Education \\ \hline
			com.seuksa.khmeredu        & Seuksa 0.2 & 05A0858F83 & Education \\ \hline
			com.wFashionCentral        & Fashion Central 0.1 & 05EA49ADAA  & Lifestyle \\ \hline
			com.wiOfferchinawholesale        & iOffer china wholesale  0.1 & 0603403D90 & Shopping \\ \hline
			com.wRichMamaDating       & Rich Mama Dating 3.4 & 0ECBD904 & Social \\ \hline
			com.wDigitalindia       & Digital india 2.1 & 091E97F669 & Productivity \\ \hline
			com.wFitterBooks       & Fitter Books 4.1 & 09E527B22F  & Education \\ \hline
			
		\end{tabular}%
\end{table*}


\noindent\framebox{\parbox{\dimexpr\linewidth-2\fboxsep-2\fboxrule}{%
		\textbf{Finding 5:}	\itshape \, 32\% of the apps in our dataset leak sensitive information via unencrypted communication protocols like HTTP. These URLs contain sensitive data such as device IDs, IP addresses, ad identifiers, locale information, and other sensitive data. }}

\section{Limitations}
\toolname is a dynamic instrumentation tool and relies on the instrumentation framework Frida to instrument the Android Framework and record the fingerprinting data. It inherits all the limitations of Frida, e.g., it is known to crash for the older version of Android apps \footnote{https://frida.re/docs/android/}. Besides, to navigate through various app activities, i.e., for coverage, \toolname relies on the automated Android tester Monkey~\cite{monkey}, and its coverage is limited to the activities visited by Monkey. Thus, \toolname misses the WebView-related Android components that Monkey does not explore. 

\section{Threats to Validity}
In this section, we discuss the threats to internal and external validity of our experiment.

\paragraph*{Internal Validity}
\toolname relies on existing dynamic analysis tools, and there are many automated Android testing tools. In particular, \toolname uses the Monkey tester, which might result in section bias. We choose the Monkey tester as the research community widely uses it, and official Android documents support it. Another threat is related to the selection of our dataset, i.e., whether the chosen apps favor \toolname. We mitigate this threat by selecting a large set of apps from the widely used AndroZoo dataset. Besides, we choose the most popular apps from the Google play store for manual analysis. One final threat is validating the results for the manually analyzed apps. To mitigate this threat, at least two authors of the paper independently performed the manual analysis and cross-validated the results.
\paragraph*{External Validity}
Threats to external validity relate to the generalization of our results, i.e., our results may not hold beyond the apps in our dataset. To mitigate this, we performed our study on a large set of apps from the widely accepted AndroZoo dataset and the most popular apps from the Google play store. Besides, the apps in our dataset belong to various categories, and the distribution over these categories is even. 

\section{Related Work}

Fingerprinting in browsers has been studied for a little more than a decade. To the best of our knowledge, three large-scale studies have been conducted on browser fingerprints. The first study~\cite{Panoptclick} showed how user-agents, list of plugins, and fonts available on a system can be used to fingerprint mobile devices. Their results showed that 83.6 of the user-agents strings are unique, hence, susceptible to fingerprinting. They coined the term \emph{browser fingerprinting}, referring to the use of system information obtained from web clients as fingerprints. \emph{AmIUnqiue} took it a step further and identified new attributes for fingerprint such as HTML canvas elements. It also identified the most common attributes in fingerprinting for mobile devices. Oliver's thesis~\cite{oliver2018fingerprinting} showed that fingerprinting is ``quite-effective'' on mobile devices based on a preliminary investigation in susceptibility of mobile browsers towards fingerprinting. Our work is placed in the context of browsers embedded in hybrid apps. Hybrid-app browsers are customized by the developer and, in contrast to standalone browsers, users have little to no influence on its security and privacy policies. Therefore, these browsers are a fertile ground for profiling users through fingerprinting. 

In a contrasting study, \emph{HidingInTheCrowd}~\cite{browserfingerpintLargeScale} studied the evolution of browser fingerprints over time. Their study shows that the number of unique fingerprints has reduced from the previous studies --- more in the case of mobile browsers than desktop browsers. The fingerprints obtained from mobile browsers, in their study, present attributes having unique values and primarily use user-agent settings and HTML canvas elements. It conforms to Oliver's study~\cite{oliver2018fingerprinting}, where it shows that a majority of mobile fingerprints are unique due to the presence of an unique identifier. This observervation also conforms with our study, where we have also obtained fingerprints which are also unique to users and devices. As ours is the first studing fingerprinting in hybrid browsers to the best of our knowledge, it is difficult to comment on the evolution of fingerprinting in hybrid browsers.

Apart from these, earlier studies have also focussed on the sources of fingerprints. Acar and others' study~\cite{CanvasFingerprintngAcar} on fingerprinting showed the use of HTML \texttt{canvas} elements in fingerprinting. Sources of fingerprinting also includes, WebGL~\cite{Cao2017CrossBrowserFV, Mowery2012PixelP}, Web Audio API~\cite{AudioEngelhardt}, browser extensions~\cite{BrowserExtensions, BrowserExtensions2, BrowserExtensions3}, and CSS querying~\cite{CSSFingerprinting}, among many others. Therefore, browser fingerprinting techniques have diversified their sources keeping in pace with evolution of the web. In comparison, we have confined to features in HTTP-headers in hybrid apps in to our study. Hybrid apps do not support browser extensions, and therefore, we have not considered these in our study. Also, we did not find other sources, such as canvas elements, WebGL resources in our study and choose to ignore these features.



The paper also overlaps with studies on privacy leakage in hybrid apps. Tiwari et al.~\cite{LuDroidSCAM,LuDroid-Journal} profiled privacy information leaked through the bridge interface. Rizzo et al.~\cite{rizzo2018babelview} studied the use of code injection attacks in WebView. Lee et al.~\cite{AdLib} discovered the vulnerability of AdSDKs leaking sensitive information via \emph{loadUrl}. Mutchler~\cite{Mutchler2015ALS} conduced a large-scale study on the Android app ecosystem to detect vulnerabilities in hybrid apps. Their findings suggest that hybrid apps have at least one security vulnerability in the Android app ecosystem. Zhang~\cite{Zhang18Usenix} performed a large-scale study of Web resource manipulation in both Android and iOS WebViews. They discovered 21 apps with malicious intents such as collecting user credentials and impersonating legitimate parties. In comparison to all these works, we analyze the fingerprints obtained from the hybrid-browsers, and manually analyze the privacy-leakage thereof.

\section{Conclusion}
In this paper, we studied the fingerprints obtained in hybrid apps. To this end, we developed an instrumentation-based tool to record the user-agent strings and HTTP headers used in the webpage of the hybrid apps. Our study shows that hybrid apps are as susceptible to fingerprints as websites accessed on mobile browsers. However, the absence of mechanisms to enforce privacy policies makes it harder, if not impossible, for users to protect their privacy. Therefore, the recent advances in protecting privacy via fingerprinting do not translate into the realm of hybrid apps as the configuration remains in the hands of developers. Our study highlights the need for research into mechanisms to enforce privacy policies in hybrid apps.
%
%
%

%
%
%
\bibliographystyle{ACM-Reference-Format}
\bibliography{biblio.bib}


\begin{thebibliography}{29}


\ifx \showCODEN    \undefined \def \showCODEN     #1{\unskip}     \fi
\ifx \showDOI      \undefined \def \showDOI       #1{#1}\fi
\ifx \showISBNx    \undefined \def \showISBNx     #1{\unskip}     \fi
\ifx \showISBNxiii \undefined \def \showISBNxiii  #1{\unskip}     \fi
\ifx \showISSN     \undefined \def \showISSN      #1{\unskip}     \fi
\ifx \showLCCN     \undefined \def \showLCCN      #1{\unskip}     \fi
\ifx \shownote     \undefined \def \shownote      #1{#1}          \fi
\ifx \showarticletitle \undefined \def \showarticletitle #1{#1}   \fi
\ifx \showURL      \undefined \def \showURL       {\relax}        \fi
\providecommand\bibfield[2]{#2}
\providecommand\bibinfo[2]{#2}
\providecommand\natexlab[1]{#1}
\providecommand\showeprint[2][]{arXiv:#2}

\bibitem[tra(2014)]%
        {tracks}
 \bibinfo{year}{2014}\natexlab{}.
\newblock \bibinfo{title}{Coveryourtracks}.
\newblock \bibinfo{howpublished}{\url{https://coveryourtracks.eff.org/}}.
\newblock


\bibitem[nig(2015)]%
        {nightwatch}
 \bibinfo{year}{2015}\natexlab{}.
\newblock \bibinfo{title}{Research: Chrome For Android Reveals Phone Model and
  Build}.
\newblock
  \bibinfo{howpublished}{\url{https://wwws.nightwatchcybersecurity.com/2015/09/30/research-chrome-for-android-reveals-phone-model-and-build/}}.
\newblock


\bibitem[chr(2015)]%
        {chromiumBug}
 \bibinfo{year}{2015}\natexlab{}.
\newblock \bibinfo{title}{Webview privacy issue}.
\newblock
  \bibinfo{howpublished}{\url{https://bugs.chromium.org/p/chromium/issues/detail?id=494452}}.
\newblock


\bibitem[web(2021)]%
        {webview}
 \bibinfo{year}{2021}\natexlab{}.
\newblock \bibinfo{title}{Webview}.
\newblock
  \bibinfo{howpublished}{\url{https://developer.android.com/reference/android/webkit/WebView/}}.
\newblock


\bibitem[fri(2022)]%
        {frida}
 \bibinfo{year}{2022}\natexlab{}.
\newblock \bibinfo{title}{Dynamic instrumentation toolkit for developers,
  reverse-engineers, and security researchers}.
\newblock \bibinfo{howpublished}{\url{https://frida.re/docs/android/}}.
\newblock


\bibitem[zen(2022)]%
        {zenodo}
 \bibinfo{year}{2022}\natexlab{}.
\newblock \bibinfo{title}{Tool and dataset for fingerprinting the Android
  hybrid web apps}.
\newblock \bibinfo{howpublished}{https://doi.org/10.5281/zenodo.6779325}.
\newblock


\bibitem[Acar et~al\mbox{.}(2014)]%
        {CanvasFingerprintngAcar}
\bibfield{author}{\bibinfo{person}{Gunes Acar}, \bibinfo{person}{Christian
  Eubank}, \bibinfo{person}{Steven Englehardt}, \bibinfo{person}{Marc Juarez},
  \bibinfo{person}{Arvind Narayanan}, {and} \bibinfo{person}{Claudia Diaz}.}
  \bibinfo{year}{2014}\natexlab{}.
\newblock \showarticletitle{The Web Never Forgets: Persistent Tracking
  Mechanisms in the Wild}. In \bibinfo{booktitle}{\emph{Proceedings of the 2014
  ACM SIGSAC Conference on Computer and Communications Security}} (Scottsdale,
  Arizona, USA) \emph{(\bibinfo{series}{CCS '14})}.
  \bibinfo{publisher}{Association for Computing Machinery},
  \bibinfo{address}{New York, NY, USA}, \bibinfo{pages}{674--689}.
\newblock
\showISBNx{9781450329576}
\urldef\tempurl%
\url{https://doi.org/10.1145/2660267.2660347}
\showDOI{\tempurl}


\bibitem[Allix et~al\mbox{.}(2016)]%
        {allix2016androzoo}
\bibfield{author}{\bibinfo{person}{Kevin Allix},
  \bibinfo{person}{Tegawend{\'e}~F Bissyand{\'e}}, \bibinfo{person}{Jacques
  Klein}, {and} \bibinfo{person}{Yves Le~Traon}.}
  \bibinfo{year}{2016}\natexlab{}.
\newblock \showarticletitle{Androzoo: Collecting millions of android apps for
  the research community}. In \bibinfo{booktitle}{\emph{2016 IEEE/ACM 13th
  Working Conference on Mining Software Repositories (MSR)}}. IEEE,
  \bibinfo{pages}{468--471}.
\newblock


\bibitem[Cao et~al\mbox{.}(2017)]%
        {Cao2017CrossBrowserFV}
\bibfield{author}{\bibinfo{person}{Yinzhi Cao}, \bibinfo{person}{Song Li},
  {and} \bibinfo{person}{Erik Wijmans}.} \bibinfo{year}{2017}\natexlab{}.
\newblock \showarticletitle{(Cross-)Browser Fingerprinting via OS and Hardware
  Level Features}. In \bibinfo{booktitle}{\emph{NDSS}}.
\newblock


\bibitem[Eckersley(2010)]%
        {Panoptclick}
\bibfield{author}{\bibinfo{person}{Peter Eckersley}.}
  \bibinfo{year}{2010}\natexlab{}.
\newblock \showarticletitle{How Unique is Your Web Browser?}. In
  \bibinfo{booktitle}{\emph{Proceedings of the 10th International Conference on
  Privacy Enhancing Technologies}} (Berlin, Germany)
  \emph{(\bibinfo{series}{PETS'10})}. \bibinfo{publisher}{Springer-Verlag},
  \bibinfo{address}{Berlin, Heidelberg}, \bibinfo{pages}{1--18}.
\newblock
\showISBNx{3642145264}


\bibitem[Englehardt and Narayanan(2016)]%
        {AudioEngelhardt}
\bibfield{author}{\bibinfo{person}{Steven Englehardt} {and}
  \bibinfo{person}{Arvind Narayanan}.} \bibinfo{year}{2016}\natexlab{}.
\newblock \showarticletitle{Online Tracking: A 1-Million-Site Measurement and
  Analysis}. In \bibinfo{booktitle}{\emph{Proceedings of the 2016 ACM SIGSAC
  Conference on Computer and Communications Security}} (Vienna, Austria)
  \emph{(\bibinfo{series}{CCS '16})}. \bibinfo{publisher}{Association for
  Computing Machinery}, \bibinfo{address}{New York, NY, USA},
  \bibinfo{pages}{1388--1401}.
\newblock
\showISBNx{9781450341394}
\urldef\tempurl%
\url{https://doi.org/10.1145/2976749.2978313}
\showDOI{\tempurl}


\bibitem[G\'{o}mez-Boix et~al\mbox{.}(2018)]%
        {browserfingerpintLargeScale}
\bibfield{author}{\bibinfo{person}{Alejandro G\'{o}mez-Boix},
  \bibinfo{person}{Pierre Laperdrix}, {and} \bibinfo{person}{Benoit Baudry}.}
  \bibinfo{year}{2018}\natexlab{}.
\newblock \showarticletitle{Hiding in the Crowd: An Analysis of the
  Effectiveness of Browser Fingerprinting at Large Scale}. In
  \bibinfo{booktitle}{\emph{Proceedings of the 2018 World Wide Web Conference}}
  (Lyon, France) \emph{(\bibinfo{series}{WWW '18})}.
  \bibinfo{publisher}{International World Wide Web Conferences Steering
  Committee}, \bibinfo{address}{Republic and Canton of Geneva, CHE},
  \bibinfo{pages}{309--318}.
\newblock
\showISBNx{9781450356398}
\urldef\tempurl%
\url{https://doi.org/10.1145/3178876.3186097}
\showDOI{\tempurl}


\bibitem[Google(2022a)]%
        {dumm}
\bibfield{author}{\bibinfo{person}{Google}.} \bibinfo{year}{2022}\natexlab{a}.
\newblock \bibinfo{title}{Chromium WebView Browser}.
\newblock
  \bibinfo{howpublished}{\url{https://developer.chrome.com/docs/multidevice/webview/}}.
\newblock


\bibitem[Google(2022b)]%
        {monkey}
\bibfield{author}{\bibinfo{person}{Google}.} \bibinfo{year}{2022}\natexlab{b}.
\newblock \bibinfo{title}{Monkey Tester}.
\newblock
  \bibinfo{howpublished}{\url{https://developer.android.com/studio/test/other-testing-tools/monkey}}.
\newblock


\bibitem[Gulyas et~al\mbox{.}(2018)]%
        {BrowserExtensions3}
\bibfield{author}{\bibinfo{person}{Gabor~Gyorgy Gulyas},
  \bibinfo{person}{Doliere~Francis Some}, \bibinfo{person}{Nataliia Bielova},
  {and} \bibinfo{person}{Claude Castelluccia}.}
  \bibinfo{year}{2018}\natexlab{}.
\newblock \showarticletitle{To Extend or Not to Extend: On the Uniqueness of
  Browser Extensions and Web Logins}. In \bibinfo{booktitle}{\emph{Proceedings
  of the 2018 Workshop on Privacy in the Electronic Society}} (Toronto, Canada)
  \emph{(\bibinfo{series}{WPES'18})}. \bibinfo{publisher}{Association for
  Computing Machinery}, \bibinfo{address}{New York, NY, USA},
  \bibinfo{pages}{14--27}.
\newblock
\showISBNx{9781450359894}
\urldef\tempurl%
\url{https://doi.org/10.1145/3267323.3268959}
\showDOI{\tempurl}


\bibitem[(IETF)(2022)]%
        {rfc9110}
\bibfield{author}{\bibinfo{person}{Internet Engineering Task~Force (IETF)}.}
  \bibinfo{year}{2022}\natexlab{}.
\newblock \bibinfo{title}{RFC 9110: HTTP Semantics}.
\newblock
  \bibinfo{howpublished}{\url{https://www.rfc-editor.org/rfc/rfc9110\#name-user-agent}}.
\newblock


\bibitem[Laperdrix et~al\mbox{.}(2020)]%
        {bsurvey}
\bibfield{author}{\bibinfo{person}{Pierre Laperdrix}, \bibinfo{person}{Nataliia
  Bielova}, \bibinfo{person}{Benoit Baudry}, {and} \bibinfo{person}{Gildas
  Avoine}.} \bibinfo{year}{2020}\natexlab{}.
\newblock \showarticletitle{Browser Fingerprinting: A Survey}.
\newblock \bibinfo{journal}{\emph{ACM Trans. Web}} \bibinfo{volume}{14},
  \bibinfo{number}{2}, Article \bibinfo{articleno}{8} (\bibinfo{date}{apr}
  \bibinfo{year}{2020}), \bibinfo{numpages}{33}~pages.
\newblock
\showISSN{1559-1131}
\urldef\tempurl%
\url{https://doi.org/10.1145/3386040}
\showDOI{\tempurl}


\bibitem[Laperdrix et~al\mbox{.}(2016)]%
        {AmIUnique}
\bibfield{author}{\bibinfo{person}{Pierre Laperdrix}, \bibinfo{person}{Walter
  Rudametkin}, {and} \bibinfo{person}{Benoit Baudry}.}
  \bibinfo{year}{2016}\natexlab{}.
\newblock \showarticletitle{Beauty and the Beast: Diverting Modern Web Browsers
  to Build Unique Browser Fingerprints}. In \bibinfo{booktitle}{\emph{2016 IEEE
  Symposium on Security and Privacy (SP)}}. \bibinfo{pages}{878--894}.
\newblock
\urldef\tempurl%
\url{https://doi.org/10.1109/SP.2016.57}
\showDOI{\tempurl}


\bibitem[Lee and Ryu(2019)]%
        {AdLib}
\bibfield{author}{\bibinfo{person}{Sungho Lee} {and} \bibinfo{person}{Sukyoung
  Ryu}.} \bibinfo{year}{2019}\natexlab{}.
\newblock \bibinfo{booktitle}{\emph{Adlib: Analyzer for Mobile Ad Platform
  Libraries}}.
\newblock \bibinfo{publisher}{Association for Computing Machinery},
  \bibinfo{address}{New York, NY, USA}, \bibinfo{pages}{262--272}.
\newblock
\showISBNx{9781450362245}
\urldef\tempurl%
\url{https://doi.org/10.1145/3293882.3330562}
\showURL{%
\tempurl}


\bibitem[Mowery and Shacham(2012)]%
        {Mowery2012PixelP}
\bibfield{author}{\bibinfo{person}{Keaton Mowery} {and} \bibinfo{person}{Hovav
  Shacham}.} \bibinfo{year}{2012}\natexlab{}.
\newblock \showarticletitle{Pixel Perfect : Fingerprinting Canvas in HTML 5}.
\newblock


\bibitem[Mutchler et~al\mbox{.}(2015)]%
        {Mutchler2015ALS}
\bibfield{author}{\bibinfo{person}{Patrick Mutchler}, \bibinfo{person}{Adam
  Doup{\'e}}, \bibinfo{person}{John~C. Mitchell}, \bibinfo{person}{Christopher
  Kruegel}, {and} \bibinfo{person}{Giovanni Vigna}.}
  \bibinfo{year}{2015}\natexlab{}.
\newblock \showarticletitle{A Large-Scale Study of Mobile Web App Security}.
\newblock


\bibitem[Oliver(2018)]%
        {oliver2018fingerprinting}
\bibfield{author}{\bibinfo{person}{John Oliver}.}
  \bibinfo{year}{2018}\natexlab{}.
\newblock \emph{\bibinfo{title}{Fingerprinting the Mobile Web}}.
\newblock \bibinfo{thesistype}{Ph.\,D. Dissertation}. \bibinfo{school}{Master
  Thesis. London, UK: Imperial College London}.
\newblock


\bibitem[Rizzo et~al\mbox{.}(2018)]%
        {rizzo2018babelview}
\bibfield{author}{\bibinfo{person}{Claudio Rizzo}, \bibinfo{person}{Lorenzo
  Cavallaro}, {and} \bibinfo{person}{Johannes Kinder}.}
  \bibinfo{year}{2018}\natexlab{}.
\newblock \showarticletitle{Babelview: Evaluating the impact of code injection
  attacks in mobile webviews}. In \bibinfo{booktitle}{\emph{International
  Symposium on Research in Attacks, Intrusions, and Defenses}}. Springer,
  \bibinfo{pages}{25--46}.
\newblock


\bibitem[Sanchez-Rola et~al\mbox{.}(2017)]%
        {BrowserExtensions2}
\bibfield{author}{\bibinfo{person}{Iskander Sanchez-Rola},
  \bibinfo{person}{Igor Santos}, {and} \bibinfo{person}{Davide Balzarotti}.}
  \bibinfo{year}{2017}\natexlab{}.
\newblock \showarticletitle{Extension Breakdown: Security Analysis of Browsers
  Extension Resources Control Policies}. In \bibinfo{booktitle}{\emph{26th
  USENIX Security Symposium (USENIX Security 17)}}. \bibinfo{publisher}{USENIX
  Association}, \bibinfo{address}{Vancouver, BC}, \bibinfo{pages}{679--694}.
\newblock
\showISBNx{978-1-931971-40-9}
\urldef\tempurl%
\url{https://www.usenix.org/conference/usenixsecurity17/technical-sessions/presentation/sanchez-rola}
\showURL{%
\tempurl}


\bibitem[Starov and Nikiforakis(2017)]%
        {BrowserExtensions}
\bibfield{author}{\bibinfo{person}{Oleksii Starov} {and} \bibinfo{person}{Nick
  Nikiforakis}.} \bibinfo{year}{2017}\natexlab{}.
\newblock \showarticletitle{XHOUND: Quantifying the Fingerprintability of
  Browser Extensions}. In \bibinfo{booktitle}{\emph{2017 IEEE Symposium on
  Security and Privacy (SP)}}. \bibinfo{pages}{941--956}.
\newblock
\urldef\tempurl%
\url{https://doi.org/10.1109/SP.2017.18}
\showDOI{\tempurl}


\bibitem[Takei et~al\mbox{.}(2015)]%
        {CSSFingerprinting}
\bibfield{author}{\bibinfo{person}{Naoki Takei}, \bibinfo{person}{Takamichi
  Saito}, \bibinfo{person}{Ko Takasu}, {and} \bibinfo{person}{Tomotaka
  Yamada}.} \bibinfo{year}{2015}\natexlab{}.
\newblock \showarticletitle{Web Browser Fingerprinting Using Only Cascading
  Style Sheets}. In \bibinfo{booktitle}{\emph{2015 10th International
  Conference on Broadband and Wireless Computing, Communication and
  Applications (BWCCA)}}. \bibinfo{pages}{57--63}.
\newblock
\urldef\tempurl%
\url{https://doi.org/10.1109/BWCCA.2015.105}
\showDOI{\tempurl}


\bibitem[Tiwari et~al\mbox{.}(2019)]%
        {LuDroidSCAM}
\bibfield{author}{\bibinfo{person}{A. Tiwari}, \bibinfo{person}{J. Prakash},
  \bibinfo{person}{S. GroB}, {and} \bibinfo{person}{C. Hammer}.}
  \bibinfo{year}{2019}\natexlab{}.
\newblock \showarticletitle{LUDroid: A Large Scale Analysis of Android -- Web
  Hybridization}. In \bibinfo{booktitle}{\emph{2019 IEEE 19th International
  Working Conference on Source Code Analysis and Manipulation (SCAM)}}.
  \bibinfo{publisher}{IEEE Computer Society}, \bibinfo{address}{Los Alamitos,
  CA, USA}, \bibinfo{pages}{256--267}.
\newblock
\urldef\tempurl%
\url{https://doi.org/10.1109/SCAM.2019.00036}
\showDOI{\tempurl}


\bibitem[Tiwari et~al\mbox{.}(2020)]%
        {LuDroid-Journal}
\bibfield{author}{\bibinfo{person}{Abhishek Tiwari}, \bibinfo{person}{Jyoti
  Prakash}, \bibinfo{person}{Sascha Gro{\ss}}, {and} \bibinfo{person}{Christian
  Hammer}.} \bibinfo{year}{2020}\natexlab{}.
\newblock \showarticletitle{A Large Scale Analysis of Android --- Web
  Hybridization}.
\newblock \bibinfo{journal}{\emph{Journal of Systems and Software}}
  \bibinfo{volume}{170} (\bibinfo{year}{2020}), \bibinfo{pages}{110775}.
\newblock
\showISSN{0164-1212}
\urldef\tempurl%
\url{https://doi.org/10.1016/j.jss.2020.110775}
\showDOI{\tempurl}


\bibitem[Zhang et~al\mbox{.}(2018)]%
        {Zhang18Usenix}
\bibfield{author}{\bibinfo{person}{Xiaohan Zhang}, \bibinfo{person}{Yuan
  Zhang}, \bibinfo{person}{Qianqian Mo}, \bibinfo{person}{Hao Xia},
  \bibinfo{person}{Zhemin Yang}, \bibinfo{person}{Min Yang},
  \bibinfo{person}{Xiaofeng Wang}, \bibinfo{person}{Long Lu}, {and}
  \bibinfo{person}{Haixin Duan}.} \bibinfo{year}{2018}\natexlab{}.
\newblock \showarticletitle{An Empirical Study of Web Resource Manipulation in
  Real-World Mobile Applications}. In \bibinfo{booktitle}{\emph{Proceedings of
  the 27th USENIX Conference on Security Symposium}} (Baltimore, MD, USA)
  \emph{(\bibinfo{series}{SEC'18})}. \bibinfo{publisher}{USENIX Association},
  \bibinfo{address}{USA}, \bibinfo{pages}{1183–1198}.
\newblock
\showISBNx{9781931971461}


\end{thebibliography}

\end{document}